\documentclass[journal=cgdefu,manuscript=communication]{achemso}

\usepackage[version=3]{mhchem} % Formula subscripts using \ce{}

\author{Shinya Yamada}
\affiliation[Kyushu University]
{Department of Electronics, Kyushu University, 744 Motooka, Fukuoka 819-0395, Japan}
\author{Kohei Tanikawa}
\affiliation[Kyushu University]
{Department of Electronics, Kyushu University, 744 Motooka, Fukuoka 819-0395, Japan}
\author{Masanobu Miyao}
\affiliation[Kyushu University]
{Department of Electronics, Kyushu University, 744 Motooka, Fukuoka 819-0395, Japan}
\author{Kohei Hamaya}
\email{hamaya@ed.kyushu-u.ac.jp}
\affiliation[Kyushu University]
{Department of Electronics, Kyushu University, 744 Motooka, Fukuoka 819-0395, Japan}

\title{Atomically controlled epitaxial growth of single-crystalline germanium films on a metallic silicide}

\begin{document}
%%%%%%%%%%%%%%%%%%%%%%%%%%%%%%%%%%%%%%%%%%%%%%%%%%%%%%%%%%%%%%%%%%%%%
%% The manuscript does not need to include \maketitle, which is
%% executed automatically.  The document should begin with an
%% abstract, if appropriate.  If one is given and should not be, the
%% contents will be gobbled.
%%%%%%%%%%%%%%%%%%%%%%%%%%%%%%%%%%%%%%%%%%%%%%%%%%%%%%%%%%%%%%%%%%%%%
\begin{abstract}
We demonstrate high-quality epitaxial germanium (Ge) films on a metallic silicide, Fe$_{3}$Si, grown directly on a Ge(111) substrate. Using molecular beam epitaxy techniques, we can obtain an artificially controlled arrangement of silicon (Si) or iron (Fe) atoms at the surface on Fe$_{3}$Si(111). The Si-terminated Fe$_{3}$Si(111) surface enables us to grow two-dimensional epitaxial Ge films whereas the Fe-terminated one causes the three-dimensional epitaxial growth of Ge films. The high-quality Ge grown on the Si-terminated surface has almost no strain, meaning that the Ge films are not grown on the low-temperature-grown Si buffer layer but on the lattice matched metallic Fe$_{3}$Si. This study will open a new way for vertical-type Ge-channel transistors with metallic source/drain contacts.

\end{abstract}

%\section{INTRODUCTION}
%\section{EXPERIMENTAL SECTION}
%\section{RESULTS AND DISCUSSION}
%\section{CONCLUSION}

\clearpage
Since germanium (Ge) has relatively high electron and hole mobility compared with silicon (Si)\cite{Tezuka,Miyao}, significant progresses with gate-stacking\cite{Ritenour,Lee,Hirayama}, source-drain\cite{Nakaharai,Tezuka,Nishimura2,Dimoulas,Yamane,Kasahara,Takeuchi}, and thin-film channel\cite{Miyao,Ikeda,Myronov,Nakamura,Hoshi} technologies have so far been reported for Ge-based metal-oxide-semiconductor filed-effect transistors (MOSFETs). Whereas the use of the Ge channel can be good solution in order to keep the development of complementary metal-oxide-semiconductor (CMOS) technologies, critical limitations still remain from the viewpoint of scalability because of their lateral device structures fabricated by the lithography processes.

On the other hand, vertical-type device structures have also been proposed. To achieve ultrahigh-density nanoscale devices for three-dimensional integrated circuits, oriented epitaxial Ge nanowires were explored.\cite{Adhikari,Kamins,Fukata} In particular, growth and impurity doping technologies compatible with Si-based CMOS technologies have been well investigated.\cite{Adhikari,Kamins,Fukata} However, there are still many technological issues such as precise controls of crystal orientation and of the doping profile for the Ge nanowires. That is, it is difficult for vertical-type Ge nanodevice structures to achieve low-resistance contacts by using an impurity doping technique. Consequently, the artificially fabricated metal/Ge/metal structures without impurity doping will be desirable for next-generation devices.

In general, there is a large difference in crystallization energy between element semiconductors such as Si or Ge and metals because of the difference in bonding energy between covalent and metallic bonds. Also, Si or Ge essentially requires relatively high thermal energy to crystallize, whereas metals can crystallize with a relatively low one. As a result, if one tries to form high-quality Si or Ge on a metal in conventional growth conditions, the formation of other compounds such as silicide or germanide cannot be prevented. Up to now, to realize Si-based three-dimensional integrated devices, the crystal growth of Si/metal/Si vertical double heterostructures has already been studied.\cite{Saitoh,Bean,Ohsima} To overcome the above issues in the crystallization of Si, complicated methods including the combination of molecular beam epitaxy (MBE), solid phase epitaxy (SPE), and several annealing techniques were explored intensively.\cite{Saitoh,Bean,Ohsima} 

Considering an achievement of metal/Ge/metal structures, we should first form high-quality single-crystalline Ge films on a metal. In this paper, we demonstrate a high-quality Ge film on a metallic silicide grown on Ge(111). Actually, this structure is Ge/metal/Ge as well as Si/metal/Si reported so far. However, this structure can be achieved only by using an MBE technique with an artificially controlled arrangement of the surface atoms on the metallic silicide. This study will open a new way for vertical-type Ge-channel transistors with metallic source/drain contacts.

To realize epitaxial growth of Ge films on a metal, we focus on one of the silicide compounds, Fe$_{3}$Si,\cite{Hines,Niculescu,Bansil,Ionescu} as a metallic material in this study. The positive reasons are as follows. First, we have so far developed single-crystalline Fe$_{3}$Si on Si or Ge,\cite{Hamaya,Hamaya2} indicating the low electrical resistivity and good compatibility for the SiGe technologies. Second, we have well understood the mechanism of the crystal growth of high-quality Fe$_{3}$Si films on Ge.\cite{Hamaya2} Since the formed Fe$_{3}$Si/Ge heterointerface is atomically flat, the surface of the grown Fe$_{3}$Si is also atomically smooth, leading to a good condition for the growth of Ge films. Finally, the atomically flat Fe$_{3}$Si/Ge junctions have positive possibilities to solve the Fermi-level pinning problems at metal/Ge interfaces.\cite{Yamane,Kasahara} These are based on the controlled molecular beam epitaxy (MBE) techniques for a metal/Ge interface.
\begin{figure}
\includegraphics[width=8cm]{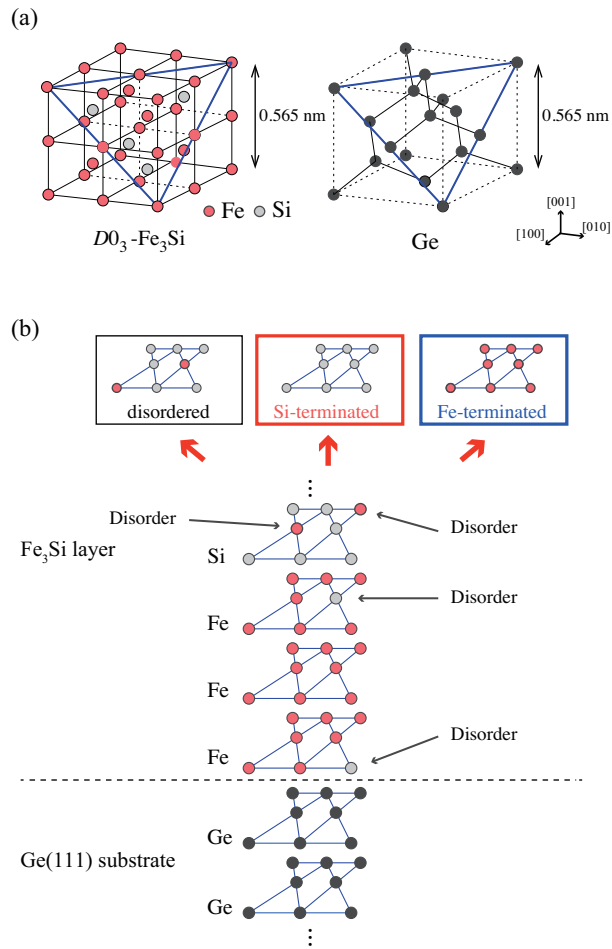}
\caption{(a) The crystal structures of $D0_{3}$-ordered Fe$_{3}$Si and Ge. (b) Schematic diagrams of the atomic arrangements of our as-grown Fe$_{3}$Si films on Ge(111) (lower) and the concept with an artificially controlled arrangement of the surface atoms on Fe$_{3}$Si (upper).}
\end{figure}

The above features arise from a special condition between Fe$_{3}$Si and Ge. Figure 1a illustrates crystal structures of the ideal ($D0_{3}$-ordered) Fe$_{3}$Si\cite{Niculescu} and Ge, which are bcc and diamond structures, respectively, where the lattice mismatch between Fe$_{3}$Si (0.565 nm) and Ge (0.565 nm) is almost zero. When we look at the atomic arrangement at the (111) plane, the ideal Fe$_{3}$Si has a periodical stacking structure consisting of three Fe layers and one Si layer as shown in Figure 1b. Fortunately, the atomic arrangements between Fe$_{3}$Si(111) and Ge(111) are well matched. By utilizing the special conditions, we have demonstrated the two-dimensional epitaxial growth of Fe$_{3}$Si films on Ge(111) with extremely high-quality heterointerfaces.\cite{Hamaya,Hamaya2}
\begin{figure}
\includegraphics[width=8cm]{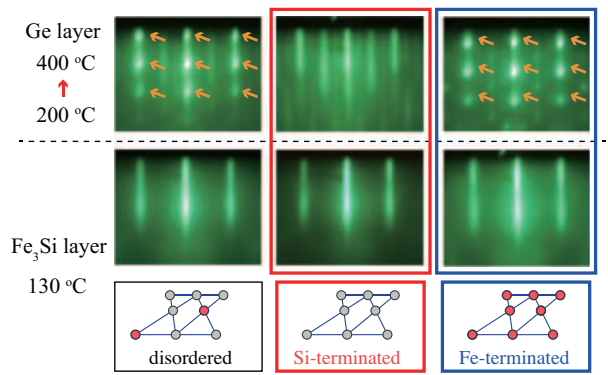}
\caption{RHEED patterns for the grown Ge films (upper) and the various surfaces of the Fe$_{3}$Si films (lower), observed along [$\overline{2}$11] azimuth.}
\end{figure}

Using the surface of the Fe$_{3}$Si/Ge(111) structure, we explore epitaxial growth of Ge films. The following is the detailed procedure for the fabrication of the Ge($\sim$100 nm)/Fe$_{3}$Si($\sim$25 nm)/Ge(111) heterostructures by using MBE techniques. Prior to the growth, we chemically cleaned nondoped Ge(111) substrates ($\rho$ $\sim$ 40 $\Omega$cm, sample size: 2 $\times$ 2 cm$^{2}$) using 1 \% HF solution to remove contamination and native oxide from the surface. The cleaned substrates were loaded immediately into an ultra high vacuum chamber with a base pressure of $\sim$ 10$^{-7}$ Pa. After the heat treatment at 550 $^{\circ}$C for 20 min, the substrate temperature was reduced down to 130 $^{\circ}$C.\cite{Hamaya,Hamaya2} After a reflection high energy electron diffraction (RHEED) pattern of the surface of the Ge(111) substrate showed it to be atomically smooth, we grew Fe$_{3}$Si films directly on the Ge(111) substrate by coevaporating Fe and Si using Knudsen cells.\cite{Hamaya} The deposition rate of Fe and Si is 2.1 nm/min and 1.2 nm/min, respectively. In-situ RHEED patterns of the Fe$_{3}$Si layers clearly exhibited symmetrical streaks, indicating good two-dimensional epitaxial growth (see the lower left RHEED pattern in Figure 2). Although the $D0_{3}$-ordered structure should have an Fe or a Si atomic layer on the top of Fe$_{3}$Si, the actual top layer of the grown Fe$_{3}$Si consists of the mixed layer with Fe and Si atoms because of some structural disorder (see upper left in Figure 1b).\cite{Hamaya2} However, an atomically smooth surface is guaranteed even for the mixed layer as shown in the lower left in Figure 2 (disordered surface). As a preliminary experiment, we try to grow Ge films (deposition rate: 0.3 nm/min) on the top of the disordered Fe$_{3}$Si with increasing the growth temperature from 200 to 400 $^{\circ}$C.\cite{Ueda} The RHEED pattern during the growth is weekly spotty [see arrows in Figure 2 (upper left)], indicating that the actual top layer of the grown Fe$_{3}$Si, consisting of the mixed layer with Fe and Si, cannot defend the three-dimensional epitaxial growth of Ge films.

Considering the above preliminary data, for the growth of high-quality Ge films on a metal, we suggest a new concept with an artificially controlled arrangement of the surface atoms on Fe$_{3}$Si. Since there is a Si atomic layer in the ideal ($D0_{3}$-ordered) Fe$_{3}$Si at the (111) plane, we can artificially form a Si(111) atomic layer on the disordered surface of the grown Fe$_{3}$Si(111) by precisely controlling the evaporation of Si atoms. Fortunately, even if the surface of the Fe$_{3}$Si layers is terminated with a few Si atomic layers, the RHEED pattern can still show streaks as displayed in the lower center of Figure 2. We note that there is almost no difference in the RHEED patterns between disordered and Si-terminated surfaces. We hereafter define the atomically smooth surface shown as the Si-terminated surface. When we formed a 2-nm-thick Si layer (six Si atomic layers) on Fe$_{3}$Si(111) at several temperatures from 130 to 350 $^{\circ}$C, the RHEED pattern darkened and disappeared, indicating that we could not crystallize the Si layers on Fe$_{3}$Si(111). Thus, we have already confirmed that there is a large difference in the surface quality between the Si-terminated surface and 2-nm-thick Si layer on Fe$_{3}$Si(111). Using the Si-terminated surface, we grow Ge layers on Fe$_{3}$Si. Interestingly, we can demonstrate two-dimensional epitaxial growth of the Ge layer (see the upper center in Figure 2) even with the same conditions shown in the above experiment. During the growth, we hardly observed the change in the RHEED pattern. As a result, a 100-nm-thick Ge epilayer can be grown even on a metallic silicide. In order to confirm the effect of the Si termination, we also investigate the growth for the Fe-terminated surface, where the surface of the Fe$_{3}$Si layers is terminated with a few Fe atomic layers as shown in the upper right panel in Figure 1b. Although the Fe-terminated surface is also atomically flat, we cannot demonstrate two-dimensional epitaxially grown Ge layer (see the right RHEED patterns in Figure 2). Note that these features were well reproduced. Therefore, we conclude that the Si termination of the disordered Fe$_{3}$Si layers is very important to obtain the high-quality Ge films on Fe$_{3}$Si(111).

To examine the structural characteristics of the grown Ge films on Fe$_{3}$Si, we observed cross-sectional transmission electron micrographs (TEM) and nanobeam electron diffraction (NED) patterns. In Figure 3b, we can see a high-quality Ge layer grown uniformly on Fe$_{3}$Si. The lattice images of the grown Ge layer are very clear. Although the Ge layer was grown directly on Fe$_{3}$Si, Fe was not detected surprisingly in the Ge layer from the energy dispersive X-ray spectroscopy (EDX) measurement, indicating no atomic interdiffusion between Ge and Fe$_{3}$Si. Also, the NED pattern of the grown Ge layer (point \#1 in Figure 3a) is almost equivalent to that of the Ge substrate (point \#2 in Figure 3a). From these results, we can conclude that a high-quality single-crystalline Ge film was obtained even on a metal. Moreover, a high-resolution image of the Ge/Fe$_{3}$Si junction clearly shows an atomically smooth heterointerface (Figure 3c). It is just as the artificially controlled heteroepitaxy of the next-generation semiconductor, Ge. We also show the surface morphology of the grown Ge layer by dynamic force mode atomic force microscopy (AFM) in Figure 3d. Since a very smooth surface with rms roughness of $\sim$ 0.41 nm can be obtained, the surface of the grown Ge layer is also a good condition for the growth of the metals such as Fe$_{3}$Si, easily enabling us to achieve metallic silicide/Ge/metallic silicide vertical structures.
\begin{figure*}
\includegraphics[width=14cm]{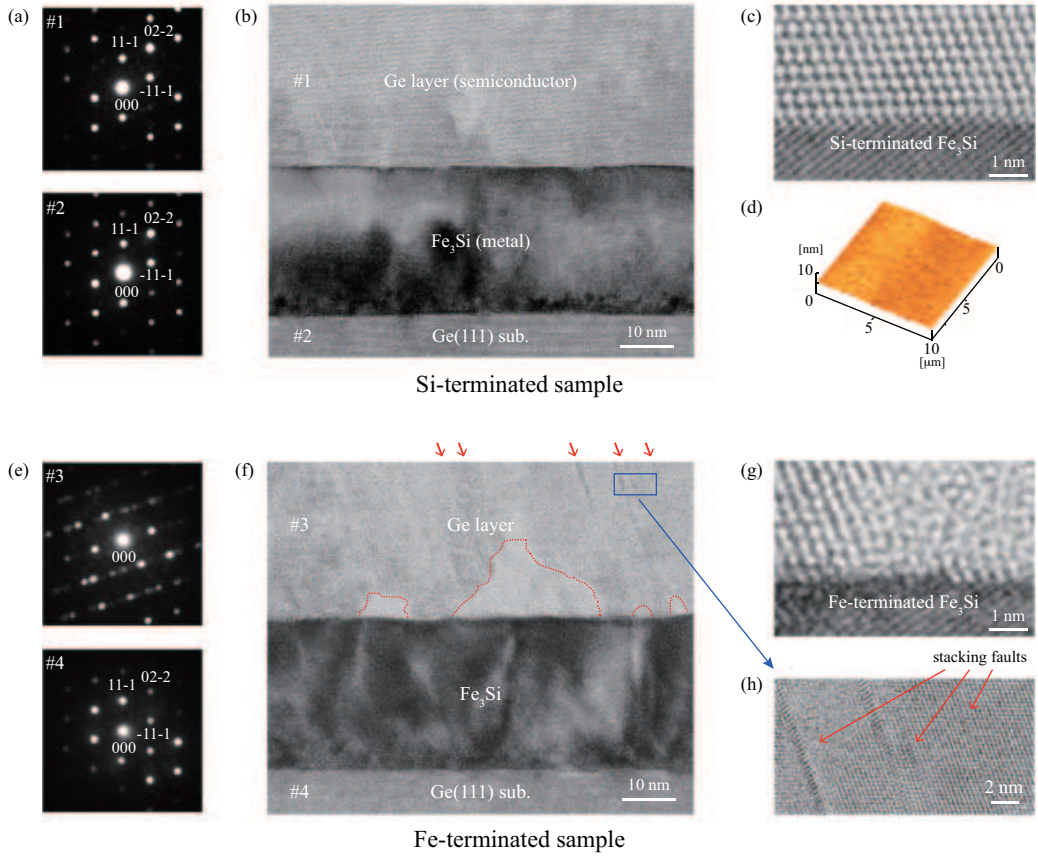}
\caption{Cross-sectional TEM images of the epitaxial Ge/metallic Fe$_{3}$Si/Ge(111) heterostructure for the growth on the Si-terminated Fe$_{3}$Si (b,c) and those on the Fe-terminated Fe$_{3}$Si (f-h). NED patterns of the epitaxial Ge film and the Ge substrate for (a) Si- and (e) Fe-terminated samples. The zone axis of the incident electron beam is parallel to [1$\overline{1}$0] direction. (d) AFM image of the surface of the Ge film grown on the Si-terminated Fe$_{3}$Si.}
\end{figure*}   

On the other hand, there are lots of stacking faults in the Ge layer on the Fe-terminated surface (Figure 3f). NED pattern measured at point \#3 also indicates the existence of stacking faults in the Ge layer. Observing the interface structures, we can clearly see amorphous phases near the interface between Ge and Fe$_{3}$Si(111), as shown in Figure 3g (also see the areas marked by red dashed lines in Figure 3f). An enlarged TEM image of the stacking faults in the Ge layer is shown in Figure 3h. We can confirm that the stacking faults in the Ge layer occur at the position near the amorphous phases. Interestingly, Fe was also not detected in the Ge layer from the EDX measurement, indicating no interdiffusion between Ge and Fe$_{3}$Si even on the Fe-terminated Fe$_{3}$Si. From the results obtained, we found that a formation of some germanide compounds (Fe-Ge) was not induced when we grew Ge films on the Fe$_{3}$Si(111) surface at a low growth temperature ($\sim$200 $^{\circ}$C). On the contrary, compared to the Si-terminated surface, some parts of Ge films were not sufficiently crystallized on the Fe-terminated surface. We can speculate that, compared to the Fe-terminated surface, the Si-terminated one reduces the crystallization temperature of Ge and/or enhances the surface migration of Ge on the Fe$_{3}$Si(111), which is caused by the difference in the bonding energy between Ge-Si bond and Ge-Fe bond. Although further study is required to clearly understand the growth mechanism of Ge on Fe$_{3}$Si, we find that the Si-terminated surface on the Fe$_{3}$Si(111) is quite effective to realize the low-temperature crystallization of Ge.

A wide-area crystal orientation of the grown Ge layer was evaluated by electron backscattering diffraction (EBSD). Figure 4a,b shows a scanning electron micrograph (SEM) (sample area: 250 $\times$ 250 $\mu$m$^{2}$) and an EBSD (sample area: 50 $\times$ 50 $\mu$m$^{2}$) image, respectively, of the Ge layer grown on the Si-terminated Fe$_{3}$Si. The surface structure of the grown Ge layer is almost uniform and the crystal orientation parallel to the growth direction is kept in the <111> direction over the entire measured area, indicating that a wide-area single-crystalline Ge(111) film was achieved even on a metal.
\begin{figure}
\includegraphics[width=8cm]{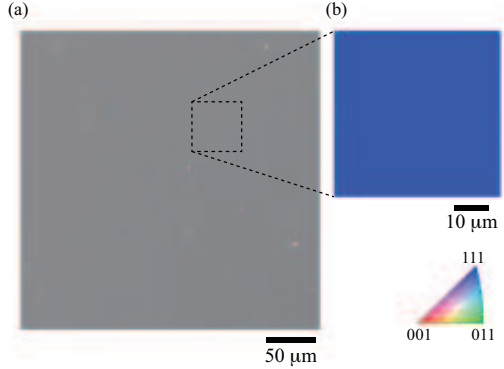}
\caption{(a) SEM and (b) EBSD images of the Ge film grown on the Si-terminated Fe$_{3}$Si.}
\end{figure} 

To further evaluate the lattice strain in the Ge layer grown at the (111) plane, we measured microprobe Raman spectra (spot size: $\sim$1 $\mu$m$\phi$, excitation laser wavelength: 532 nm, effective resolution: $\sim$0.1 cm$^{-1}$) of the 100-nm-thick Ge layer grown on Fe$_{3}$Si. Figure 5 displays a room-temperature Raman spectrum of the Ge layer grown on the Si-terminated Fe$_{3}$Si, together with that of the homoepitaxial Ge layer grown on Ge(111) at 400 $^{\circ}$C. For the Ge layer grown on the Si-terminated surface, a sharp peak originating from Ge-Ge bonding is clearly observed at $\sim$302 cm$^{-1}$. The peak position of the homoepitaxial Ge layer is also $\sim$302 cm$^{-1}$. Comparing these two data, we can recognize that there is almost no lattice strain even for the Ge layer grown on Fe$_{3}$Si. By estimating the lattice spacing from the NED patterns in Figure 3a, we also confirmed that the lattice constant of the Ge layer grown on the Si-terminated Fe$_{3}$Si is almost equal to that of bulk Ge. These features exactly support that there is no lattice mismatch between Fe$_{3}$Si and Ge, and the Si-terminated surface on Fe$_{3}$Si is not a low-temperature-grown Si buffer layer on Fe$_{3}$Si(111) but a part of a Si atomic layer in Fe$_3$Si(111). Hence, the well-matched atomic arrangement between the Si atomic layer in Fe$_{3}$Si and Ge contributes largely to the quality of the grown Ge layer.
\begin{figure}
\includegraphics[width=8cm]{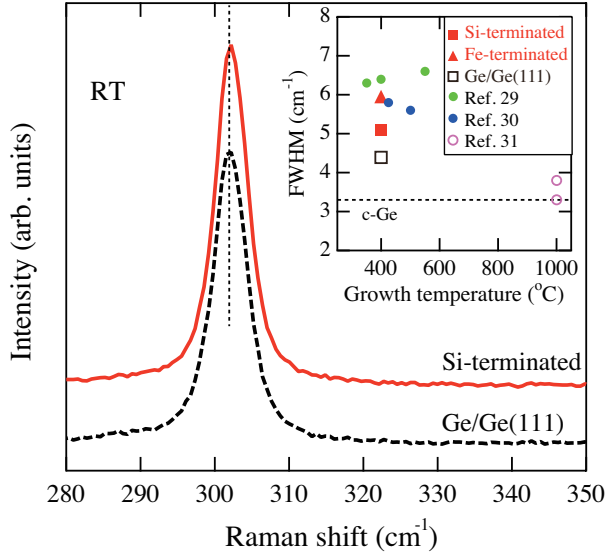}
\caption{Raman spectra of the 100-nm-thick Ge film grown on the Si-terminated Fe$_{3}$Si and the homoepitaxial Ge film grown on Ge(111) at 400 $^{\circ}$C, measured at room temperature. The inset shows the FWHM values of various Ge crystallines grown by the different methods\cite{Kanno,Toko,Toko2} as a function of growth temperature.}
\end{figure}

We finally comment on the full-width at half maximum (FWHM) value of the Raman spectra. For the Ge layer grown on Si-terminated Fe$_{3}$Si, the FWHM value was $\sim$5.1 cm$^{-1}$. We plot the FWHM values for the Ge layer grown on the Si- or Fe-terminated Fe$_{3}$Si and those of various Ge crystallines grown by the different methods\cite{Kanno,Toko,Toko2} in the inset of Figure 5. Although the FWHM value for the Ge layer grown on the Si-terminated Fe$_{3}$Si is larger than those for the Ge films grown on an insulator by a liquid-phase epitaxy ($\le$ 4.0 cm$^{-1}$),\cite{Toko2} it is at least smaller than that for the Ge layer grown on the Fe-terminated Fe$_{3}$Si ($\sim$5.9 cm$^{-1}$) and those for the crystalline Ge fabricated by other low-temperature growth techniques such as metal-induced lateral crystallization (MILC)\cite{Kanno} and solid-phase crystallization.\cite{Toko} Considering these facts, we can also say that high-quality Ge layers were obtained even on a metallic silicide. Since we can easily fabricate Fe$_{3}$Si layers on the grown Ge layer,\cite{Hamaya,Hamaya2} metallic silicide/Ge/metallic silicide vertical structures can also be easily achieved. From these results, this study will open a new way for vertical-type Ge-channel transistors with metallic source/drain contacts.

In summary, we demonstrated high-quality single-crystalline Ge films on a metallic silicide, Fe$_{3}$Si, by individually developing a novel growth technique. Using molecular beam epitaxy techniques, we obtained an artificially controlled surface structure terminated with a few Si or Fe atomic layers at the Fe$_{3}$Si(111) plane. Only the Si-terminated Fe$_{3}$Si surface enabled us to grow two-dimensional epitaxial Ge films. The high-quality Ge films grown have almost no strain, meaning that the Ge films are not grown on the low-temperature-grown Si buffer layer but on the well matched Fe$_{3}$Si. This technique can be applied for vertical-type Ge-channel transistors with metallic source/drain contacts.

\begin{acknowledgement}
The authors would like to thank Prof. T. Asano and Prof. T. Kimura of Kyushu University for their experimental support. This work was partly supported by Grant-in-Aid for Young Scientists (A) from The Japan Society for the Promotion of Science (JSPS) and Industrial Technology Research Grant Program from NEDO. S.Y. acknowledges JSPS Research Fellowships for Young Scientists. 
\end{acknowledgement}

%\begin{acknowledgement}
%
%Please use ``The authors thank \ldots'' rather than ``The
%authors would like to thank \ldots''.
%
%The author thanks Mats Dahlgren for version one of \textsf{achemso},
%and Donald Arseneau for the code taken from \textsf{cite} to move
%citations after punctuation. Many users have provided feedback on the
%class, which is reflected in all of the different demonstrations
%shown in this document.
%
%\end{acknowledgement}

%\begin{suppinfo}
%
%This will usually read something like: ``Experimental procedures and
%characterization data for all new compounds. The class will
%automatically add a sentence pointing to the information on-line:
%
%\end{suppinfo}

\bibliography{achemso-demo}

\clearpage
\begin{thebibliography}{11}
\bibitem{Tezuka}
Tezuka, T.;  Nakaharai, S.; Moriyama, Y.; Sugiyama, N.; Takagi, S.; {\it IEEE Electron Device Lett.} {\bf 2005}, {\it 26}, 243-245.
\bibitem{Miyao}
Miyao, M.; Toko, K.; Tanaka, T.; Sadoh, T.; {\it Appl. Phys. Lett.} {\bf 2009}, {\it 95}, 022115.
\bibitem{Ritenour}
Ritenour, A.; Khakifirooz, A.; Antoniadis, D. A.; Lei, R. Z.; Tsai, W.; Dimoulas, A.; Mavrou, G.; Panayiotatos, Y.; {\it Appl. Phys. Lett.} {\bf 2006}, {\it 88}, 132107.
\bibitem{Lee} 
(a) Lee, C. H.; Nishimura, T.; Tabata, T.; Wang, S. K.; Nagashio, K.; Kita, K.; Toriumi, A.; {\it IEDM Tech. Dig.} {\bf 2010}, 416-419. (b) Nishimura, T.; Lee, C. H.; Tabata, T.; Wang, S. K.; Nagashio, K; Kita, K.; Toriumi, A.; {\it Appl. Phys. Express} {\bf 2011}, {\it 4}, 064201.
\bibitem{Hirayama}
Hirayama, K.; Ueno, R.; Iwamura, Y.; Yoshino, K.; Wang, D.; Yang, H.; Nakashima, H.; {\it Jpn. J. Appl. Phys.} {\bf 2011}, {\it 50}, 04DA10.
\bibitem{Nakaharai}
Nakaharai, S.; Tezuka, T.; Sugiyama, N.; Moriyama, Y.; Takagi, S.; {\it Appl. Phys. Lett.} {\bf 2003}, {\it 83}, 3516-3518.
\bibitem{Dimoulas}
Dimoulas, A.; Tsipas, P.; Sotiropoulos, A;. Evangelou, E. K.; {\it Appl. Phys. Lett.} {\bf 2006}, {\it 89}, 252110.
\bibitem{Takeuchi}
Takeuchi, S.; Shimura, Y.; Nakatsuka, O.; Zaima, S.; Ogawa, M.; Sakai, A.; {\it Appl. Phys. Lett.} {\bf 2008}, {\it 92}, 231916.
\bibitem{Nishimura2}
Nishimura, T.; Kita, K.;  Toriumi, A.; {\it Appl. Phys. Express} {\bf 2008} {\it 1}, 051406.
\bibitem{Yamane}
Yamane, K.; Hamaya, K.; Ando, Y.; Enomoto, Y.; Yamamoto, K.; Sadoh, T.; Miyao, M.; {\it Appl. Phys. lett.} {\bf 2010}, {\it 96}, 162104.
\bibitem{Kasahara}
Kasahara, K.; Yamada, S.; Sawano, K,; Miyao, M.; Hamaya, K.; {\it Phys. Rev. B} {\bf 2011}, {\bf 84}, 205301.
\bibitem{Ikeda}
Ikeda, K.; Maeda, T.; Takagi, S.: {\it Thin Solid Films} {\bf 2006}, {\it 508}, 359-362.
\bibitem{Myronov}
Myronov, M.; Sawano, K. Shiraki. Y.; Mouri, T.; Itoh, K. M. {\it Appl. Phys. Lett.} {\bf 2007}, {\it 91}, 082108.
\bibitem{Nakamura}
Nakamura, Y.; Murayama, A.; Ichikawa, M.; {\it Crystal Growth \& Design} {\bf 2011}, {\it 11}, 3301-3305.
\bibitem{Hoshi}
Hoshi, Y.; Sawano, K.; Hamaya, K.; Miyao, M.; Shiraki, Y.; {\it Appl. Phys. Express} {\bf 2012}, {\it 5}, 015701.
\bibitem{Adhikari}
(a) Adhikari, H.; Marshall, A. F.; Chidsey, C. E. D.; Mclntyre, P. C.; {\it Nano Lett.} {\bf 2006}, {\it 6}, 318-323. (b) Woodruff, J. H.; Ratchford, J. B.; Goldthorpe, I. A.; Mclntyre, P. C.; Chidsey, C. E. D. {\it Nano Lett.} {\bf 2007}, {\it 7}, 1637-1642.
\bibitem{Kamins}
Kamins, T. I. and Li, X. and Williams, R. Stanley and Liu, X.; {\it Nano Lett.} {\bf 2004}, {\it 4}, 503-506.
\bibitem{Fukata}
Fukata, N.;Sato, K.; Mitome, M.; Bando, Y.; Sekiguchi, T.; Kirkham, M.; Hong, J.; Wang, Z. L.; Snyder, R. L.; {\it ACS Nano}, {\bf 2010}, {\it 4}, 3807-3816.
\bibitem{Saitoh}
Saitoh, S.; Ishiwara, H.; Furukawa, S.; {\it Appl. Phys. Lett.} {\bf 1980}, {\it 37}, 203-205.
\bibitem{Bean}
Bean, C. J.; Poate, M. J.; {\it Appl. Phys. Lett.} {\bf 1980}, {\it 37}, 643-646.
\bibitem{Ohsima}
(a) Ohsima, T.; Nakamura, N.; Nakagawa, K.; Miyao, M.; {\it Thin Solid Films} {\bf 1990}, {\it 184}, 275-282. (b) Miyao, M.; Nakagawa, K.; Nakamura, N.; Ohshima, T.; {\it J. Crystal Growth} {\bf 1991}, {\it 111}, 957-960.
\bibitem{Hines}
Hines, W. A.; Menotti, A. H. ; Budnick, J. I. ; Burch, T. J.; Litrenta, T.; Niculescu, V.; Raj, K.; {\it Phys. Rev. B} {\bf 1976}, {\it 13}, 4060-4068.
\bibitem{Niculescu}
Niculescu, V. A., Burch, T. J.; Budnick, J. I.; {\it J. Magn. Magn. Mater.} {\bf 1983}, {\it 39}, 223-267.
\bibitem{Bansil}
Bansil, A.; Kaprzyk, S.; Mijnarends, P. E.; Tobola, J.; {\it Phys. Rev. B} {\bf 1999}, {\it 60}, 13396-13412.
\bibitem{Ionescu}
Ionescu, A.; Vaz, C. A. F.; Trypiniotis, T.; G\"{u}rtler, C. M.; Garc\'{i}a-Miquel, H.; Bland, J. A. C.; Vickers, M. E.; Dalgliesh, R. M.; Langridge, S.; Bugoslavsky, Y.; Miyoshi, Y.; Cohen, L. F. ; Ziebeck, K. R. A.; {\it Phys. Rev. B} {\bf 2005}, {\it 71}, 094401.
\bibitem{Hamaya}
(a) Sadoh, T.; Kumano, M.; Kizuka, R.; Ueda, K.; Kenjo, A.; Miyao, M.; {\it Appl. Phys. Lett.} {\bf 2006}, {\it 89}, 182511. (b) Hamaya, K.; Ueda, K.; Kishi, Y.; Ando, Y.; Sadoh, T.; Miyao, M.; {\it Appl. Phys. Lett.} {\bf 2008}, {\it 93}, 132117.
\bibitem{Hamaya2}
(a) Hamaya, K.; Murakami, T.; Yamada, S.; Mibu K.; Miyao, M. {\it Phys. Rev. B} {\bf 2011}, {\it 83}, 144411. (b) Hamaya, K.; Ando, Y.; Sadoh, T.; Miyao, M.; {\it Jpn. J. Appl. Phys.} {\bf 2011}, {\it 50}, 010101.
\bibitem{Ueda}
Ueda, K.; Ando, Y.; Kumano, M., Sadoh, T., Maeda, Y.; Miyao, M.; {\it Appl. Surf. Sci.} {\bf 2008}, {\it 254}, 6215-6217.
\bibitem{Kanno}
Kanno, H.; Toko, K.; Sadoh, T.; Miyao, M.; {\it Appl. Phys. Lett.} {\bf 2006}, {\it 89}, 182120.
\bibitem{Toko}
Toko, K.; Nakao, I.; Sadoh, T.; Noguchi, T.; Miyao, M.; {\it Solid-State Electron} {\bf 2009}, {\bf 53}, 1159-1164.
\bibitem{Toko2}
Toko, K.; Kurosawa, M.; Yokoyama, H.; Kawabata, N.; Sakane, T.; Ohta, Y.; Tanaka, T.; Sadoh, T.; Miyao, M.; {\it Appl. Phys. Express} {\bf 2010} {\it 3}, 075603.
\end{thebibliography}

%% graphical table of contents.
%%%%%%%%%%%%%%%%%%%%%%%%%%%%%%%%%%%%%%%%%%%%%%%%%%%%%%%%%%%%%%%%%%%%%

%Some journals require a graphical entry for the Table of Contents.
%This should be laid out ``print ready'' so that the sizing of the
%Inside the \texttt{tocentry} environment, the font used is Helvetica
%8\,pt, as required by \emph{Journal of the American Chemical
%The surrounding frame is 9\,cm by 3.5\,cm, which is the maximum
%permitted for  \emph{Journal of the American Chemical Society}
%graphical table of content entries. The box will not resize if the
%content is too big: instead it will overflow the edge of the box.
%
%This box and the associated title will always be printed on a
%separate page at the end of the document.

\end{document}